\documentclass[aps,prb,twocolumn,superscriptaddress]{revtex4-1}
\usepackage{amssymb}
\usepackage[utf8]{inputenc}
\usepackage{graphicx}
\usepackage{bm,color,subfigure,amsmath}
\usepackage[normalem]{ulem}

\begin{document}
\title{Generation of magnetic skyrmion bubbles by inhomogenous spin-Hall currents}

\author{Olle~Heinonen}
\affiliation{Materials Science Division, Argonne National Laboratory, Lemont, Illinois 60439, USA}
\affiliation{Northwestern-Argonne Institute for Science and Engineering, Evanston, Illinois 60208, USA}
\author{Wanjun Jiang}
\affiliation{Materials Science Division, Argonne National Laboratory, Lemont, Illinois 60439, USA}
\author{Hamoud Somaily}
\affiliation{Materials Science Division, Argonne National Laboratory, Lemont, Illinois 60439, USA}
\affiliation{Department of Physics, Northern Illinois University, DeKalb, IL 60115}
\author{Suzanne G.E. te Velthuis}
\affiliation{Materials Science Division, Argonne National Laboratory, Lemont, Illinois 60439, USA}
\author{Axel Hoffmann}
\affiliation{Materials Science Division, Argonne National Laboratory, Lemont, Illinois 60439, USA}


\begin{abstract}
Recent experiments have shown that magnetic skyrmion bubbles can be generated and injected at room temperature in thin films\cite{Jiang2015}. Here, we demonstrate, using micromagnetic modeling, that such skyrmions can be generated by an inhomogeneous spin Hall torque in the presence of Dzyaloshinskii-Moriya interactions (DMIs). In the experimental Ta/Co$_{20}$Fe$_{60}$B$_{20}$ thin films, the DMI is rather small; nevertheless, the skyrmion bubbles are stable, or at least metastable on observational time scales.
\end{abstract}
\maketitle

\section{Introduction}

Magnetic heterostructures can support unusual spin textures that emerge because of competition between different magnetic interactions, such as interfacial, magnetostatic, exchange, and anisotropy energies\cite{HoffmannCOSSM2015}. 
In magnetic systems with broken inversion symmetry 
the Dzyaloshinskii-Moriya interaction (DMI)\cite{Dzyaloshinskii1958,Moriya} is allowed. 
The DMI favors chiral magnetization textures, and can lead to skyrmion structures. 
Skyrmions are topological
objects with a topological charge $q$ given by 
\begin{equation}
q=\int\chi\, d^2r=\frac{1}{4\pi}\int {\hat m}\cdot\left[\partial_x{\hat m}
\times\partial_y{\hat m}\right]\,d^2r,
\end{equation}
where $\chi$ is the topological (or skyrmion) charge density, and ${\hat m}$ is the three-dimensional magnetization director field in the $xy$-plane. Geometrically, $q$ is
the normalized area of the spin unit sphere that is swept out by the magnetization over its domain. Single skyrmion objects
and skyrmion crystals have indeed been observed in bulk MnSi, FeCoSi and $B20$-type Mn-based materials\cite{MuhlbauerScience2009,TokunagaNatComm2015}.
Interfacial DMI can be generated in magnetic thin films, such as CoFeB of thickness of about 1~nm with an out-of-plane effective anisotropy, grown
on top of a spin-orbit scatterer, such as Ta or Pt\cite{EmoriNatMat2013,RyuNatNano2013}. These structures can support skyrmions at room temperature and small applied magnetic fields\cite{ChenAPL2015}. In general, the skyrmions can be small\cite{KiselevJPD2011}, of a size $\approx10$~nm, or they can be larger stable or metastable bubbles stabilized by dipolar interactions in addition to the DMI. In this latter case, the DMI endows the bubble wall with a definite chirality. The skyrmions, irrespective of size, have a topological charge\cite{BraunAdvPhys2012} of $\pm1$. 
In addition, the skyrmion can have different textures independent of its charge: hedgehog or spiral skyrmions with radial or azimuthal in-plane
magnetization, respectively.


Because thin film skyrmions can be manipulated and transported, either by a spin polarized current flowing in the magnetic thin film itself, or by
spin transfer torque generated by a charge current flowing in the conducting spin-orbit scatterer, and because skyrmions are topologically protected, these systems are potentially interesting for information technology applications\cite{FertNatNano2013,HoffmannPRAppl2015}.
One problem that has to be overcome for any such realization is the controlled injection of skyrmions\cite{RommingScience2013}.
We recently demonstrated experimentally that skyrmions can be generated in a magnetic multilayer with an inhomogeneous charge current flow. The system consisted of
Ta(5~nm)/Co$_{20}$Fe$_{60}$B$_{20}$(1.1~nm)/TaO$_x$(3~nm). The multilayer was patterned into a rectangular bar 60~$\mu$m wide with a constriction at the center\cite{Jiang2015}. With a voltage
applied across the rectangular bar, an electrical current flows in the Ta layer, which induces a spin-Hall torque on the magnetization in the CoFeB. We
demonstrated that pulsed currents through the device resulted in the injection of magnetic bubbles where the device widens in the region right after the narrow
constriction. These bubbles are stable and can be moved by applying a smaller voltage, and therefore smaller current densities. By investigating the stability of the bubbles under external magnetic field and the nature of their motion induced by a current
in the Ta, we could infer that these objects are skyrmions with chiral N{\'e}el domain walls. The Kerr microscopy imaging system used to capture the objects
did not however have a sufficiently high resolution to conclusively determine the internal magnetization structure of the objects, or the exact temporal evolution during their formation.

The purpose of the present work is to demonstrate through micromagnetic modeling that an inhomogeneous spin-Hall torque such as the one
in Ref.~\onlinecite{Jiang2015} can indeed generate skyrmion bubbles, and that such bubbles are stable (or at least meta-stable on observational time scales) even though the DMI coupling is very small in CoFe-Ta systems\cite{Perez2014,EmoriPRB2014}. We identify two mechanisms that make the magnetization structure
unstable under the inhomogeneous spin-Hall torque for the relatively small value of the DMI coupling used here. In both cases, the energy supplied to the system from the spin-Hall torque through the current is large enough to overcome energy barriers from the topological protection. One is in a low-current regime in which a chiral domain wall becomes unstable at
the flaring walls of the system because of competition between spin-Hall torque, exchange interactions, and DMI, in particular the natural boundary conditions imposed on the magnetization under DMI\cite{RohartPRB2013}. The other mechanism is in a high-current regime, in which a domain wall is formed along the length of the constriction by the spin-Hall torque and injects a turbulent magnetization texture as the device widens and the spin-Hall torque
becomes inhomogeneous. Both these mechanisms result in a strongly inhomogeneous magnetization texture injected into the wide part
of the device. As the applied voltage is turned off and the concomitant spin-Hall torque vanishes, this inhomogeneous magnetization texture relaxes and can coalesce into individual chiral
bubbles, as observed experimentally. As we will show, the strength of the DMI measured experimentally, together with the other parameters that
characterize the magnetic system, puts the system in a regime where static global energy minumum skyrmion solutions are of size less than 10~nm\cite{KiselevJPD2011,RohartPRB2013}, while the skyrmion bubbles generated dynamically both in the experiment and in
the micromagnetic modeling described here are considerably larger. This dynamical generation is fundamentally different from the one studied in other parameter regimes with much larger DMI coupling, in which a spin-Hall torque can pinch off a chiral N{\'e}el stripe\cite{Lin_arxiv2015}, or convert a domain-wall pair to a skyrmion\cite{YanzhouNatCom2014}.
Our simulations suggest
that the dynamically generated skyrmions are stable, at least for the duration of all our micromagnetic solutions. 
We speculate that inhomogeneities and random pinning in the experimental system may further stabilize the bubbles; this is certainly consistent with the 
experimental observations in which magnetic bubbles move through a stick-slip motion, indicative of local pinning.

\section{Methods}
Our model system is half of the experimental system, as we are focusing on capturing the effects of the inhomogeneous current density and spin-Hall torque.
It has a narrow part of width $w$ and length $L_1$, and flares up with an opening angle of $\varphi$ degrees to a width of $W$; the flare and the wider
part have a total length $L_2$ (see Fig.~\ref{fig:figure2}).
Typically, we used $w=500$~nm, $L_1=1500$~nm, $L_2=3000$~nm, $W=3000$~nm, and $\varphi=45^\circ$, although we also did
simulations with $\varphi=90^\circ$ to ensure that the results would be obtained with a much more abrupt flare.
This system has a thin magnetic layer of thickness $t=1$~nm and a magnetization density $M_S$ of 650~emu/cm$^3$.
The out-of-plane anisotropy field is $H_a=8868$~Oe, resulting in an effective out-of-plane anisotropy field $H_{\rm eff}$ of $700$~Oe, and
the micromagnetic exchange coupling\footnote{Reported values for the micromagnetic exchange constant for CoFe alloys vary from 3.1~$\mu$erg/cm,  to about 1~$\mu$erg/cm depending on film thickness, composition, deposition conditions, substrate, and post-processing, see, for example, Refs.~\onlinecite{CoeyMMM,HeinrichUMS,VazRPP2008,ChoJMMM2013}} $A=3$~$\mu$erg/cm. 
In order to accurately capture the magnetization dynamics, we use a small dimensionless damping $\alpha=0.02$, which is in the range of 
experimentally determined values for CoFeB thin films\cite{LiuJAP2011,DevolderAPL2013}.
%
%

The CoFeB film is on top of a Ta film of conductivity $0.83$~MS/m and thin enough
that the electrostatic potential can be assumed to be constant
through the thickness of the film. 
Also, because of the higher conductivity of the Ta film than 1~nm CoFeB film and the fact that the Ta film in Ref.~\onlinecite{Jiang2015} was five times thicker than the CoFeB film, we assume that all charge current is flowing in the Ta film.
A voltage $V$
can be applied between the ends of the device; typically,
we apply potentials between 0.5~V and $5$~V. We solve Laplace's equations with boundary conditions of fixed potentials at the ends
and zero currents through the sides of the system. 
The applied voltage gives rise to a charge current density $\mathbf j$ 
which,
through the spin-orbit mediated spin-Hall effect, gives rise to an effective spin-Hall field 
$
{\mathbf H}_{\rm sh}(x,y)=H_{\rm sh}^0\left[\hat m\times
\left(\hat z\times{\hat j}\right)\right],
\label{eqn:sh_field}
$
with $H_{\rm sh}^0=\hbar/(2|e|)\theta_{\rm sh}j/(tM_S)$, where $\theta_{\rm sh}$ is the spin-Hall angle, here taken to be $10$~\%\cite{HoffmannIEEE2013}, and $\hat j$ the current density director. To obtain the dynamics of the system, we integrated the Landau-Lifshitz equation for the 3D magnetization director
$\hat m$:
\begin{equation}
\frac{d\hat m}{dt}  =  -\frac{|\gamma_e|}{1+\alpha^2}\hat m\times{\mathbf H}_{\rm eff}-\frac{|\gamma_e|\alpha}{1+\alpha^2}
\hat m\times\left[\hat m\times{\mathbf H}_{\rm eff}\right],
\label{eqn:SH_field}
\end{equation}
where $\gamma_e$ is the electron gyromagnetic constant and
the effective field ${\mathbf H}_{\rm eff}$ includes
exchange field, magnetostatic fields, anisotropic
effective field, external field, spin Hall effective field,  
and DMI effective field. The latter  is given by
$
{\mathbf H}_{DMI}=\frac{2D}{M_S}\left[\left(
\nabla\cdot\hat m\right)\hat z-\nabla m_z\right].
$
The DMI interaction $D$ in CoFeB on Ta has been estimated earlier\cite{Perez2014} to be about $0.05$~erg/cm$^2$. 
We measured the DMI interaction in our system by fitting the evolution of the magnetic domains' period with external fields, obtained by Kerr microscopy, to well-established domain models\cite{Malek1958,Gehring1993}. From these calculations, we found\cite{Somailyunpub} a DW surface energy density $\sigma_{\rm DW} =1.4$~mJ/m$^2$.  
The wall energy varies with DMI constant according to Refs.~\onlinecite{Heide2008,Woo2015}, and thus we arrived at a DMI value of 0.5 to 0.53~erg/cm$^2$,  considerably larger than that measure by Perez {\em et al.,}\cite{Perez2014} but
still rather small. 
With these parameters, we obtain a dimensionless DMI interaction 
$\kappa=\pi D/[4\sqrt{A(K_u-2\pi M_S^2)}]=D/D_c\approx0.5$, where $D_c$ is the critical DMI strength at which the chiral domain wall energy becomes zero. 

\begin{figure}
\includegraphics[width=3.5in]{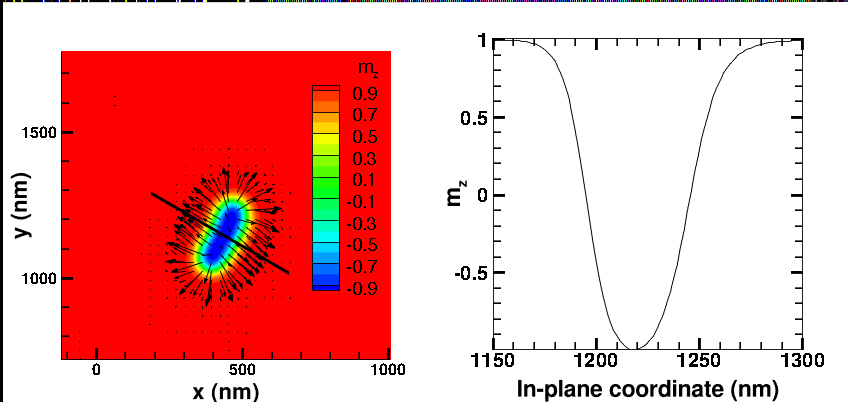}
\caption{(Color online) The left panel shows a contour plot of the out-of-plane component of the magnetization director of a skyrmion
bubble. The right panel shows the out-of-plane component of the magnetization across the line indicated in the left panel. The domain wall width estimated from $m_z=0.9$ to $m_z=-0.9$ is about 30~nm.}
\label{fig:DWwidth}
\end{figure}

The magnetic system is discretized into a mesh of dimensions $5$~nm$\times5$~nm$\times1$~nm. In comparison, the domain wall width of the skyrmion bubbles are about 30~nm, and are therefore well resolved by this mesh. (Fig. \ref{fig:DWwidth}).
We did also perform some simulations with a mesh of dimensions $2.5$~nm$\times2.5$~nm$\times1$~nm to ensure that our results
did not appreciably depend on the meshing. We note that the parallelepiped mesh used here typically gives rise to a 
staircase approximation of boundaries at an angle to the Cartesian coordinate axes. In order to avoid numerical artifacts arising from local boundary condition applied to such staircase boundaries, we smoothed the boundary conditions for the Laplace equation 
by requiring that the component of the gradient of the electrostatic potential perpendicular to the {\em average} direction of the boundary be zero (except for at corners). 
Similarly, we averaged the DMI boundary conditions along the sloped boundaries over three cells.
We did not apply any smoothing procedure to the magnetostatic fields as they are obtained from convolving bulk and surface charge densities with a slowly decaying kernel which effectively smoothens the effect of the staircase boundary.
The left panel of Fig.~\ref{fig:DWwidth} 
 shows the magnetization of a skyrmion bubble; the right panel shows the $z$-component of the 
magnetization director on a line across the magnetic skyrmion (indicated in the left panel). In this case the applied voltage was initially 5~V for a duration of 40~ns, after which the magnetization was relaxed for 80~ns with a magnetic skyrmion bubble remaining. The domain wall width estimated from $m_z=0.9$ to $m_z=-0.9$ is about 30~nm.

We apply a uniform out-of-plane field of strength $5$~Oe, and initialize the system with a domain wall
located at the beginning of the flare (see Fig.~\ref{fig:figure2} upper left panel). 
We first relax the system by integrating the
Landau-Lifshitz-Gilbert (LLG) equation with a dimensionless damping of $\alpha=0.25$ (for more details on the micromagnetic solver, see for example,
Refs.~\onlinecite{Heinonen2007,Schreiber2009}). We then apply a potential difference $V$ along the sample, solve for the charge current density, and integrate the LLG equation, now with the addition of the spin-Hall torque and with a dimensionless damping
$\alpha=0.02$ for 30 -- 40~ns. After this, we turn off the voltage (and so the charge current and spin-Hall torque) and let the system relax for 40 -- 80~ns with $\alpha=0.02$. We have not considered here the effect of a finite decay-time of the voltage pulse under the assumption that the experimental decay time is small enough that the magnetization dynamics do not respond adiabatically to turning off the voltage. The typical time steps used during the integration were kept to less than or equal to 0.5~ps, which afforded good stability.
Finite temperatures are not included in the modeling as our goal was to elucidate instability mechanisms caused by the magnetic interactions as well as the stability of generated skyrmion bubbles. Furthermore, the temperature increase due to Joule heating in Ref.~\onlinecite{Jiang2015} was estimated to less than 3~K in that work, which clearly indicates that local Joule heating does not play any significant role in the formation of the skyrmion bubbles.


\section{Results and Discussion}
\begin{figure}
\includegraphics[width=3.5in]{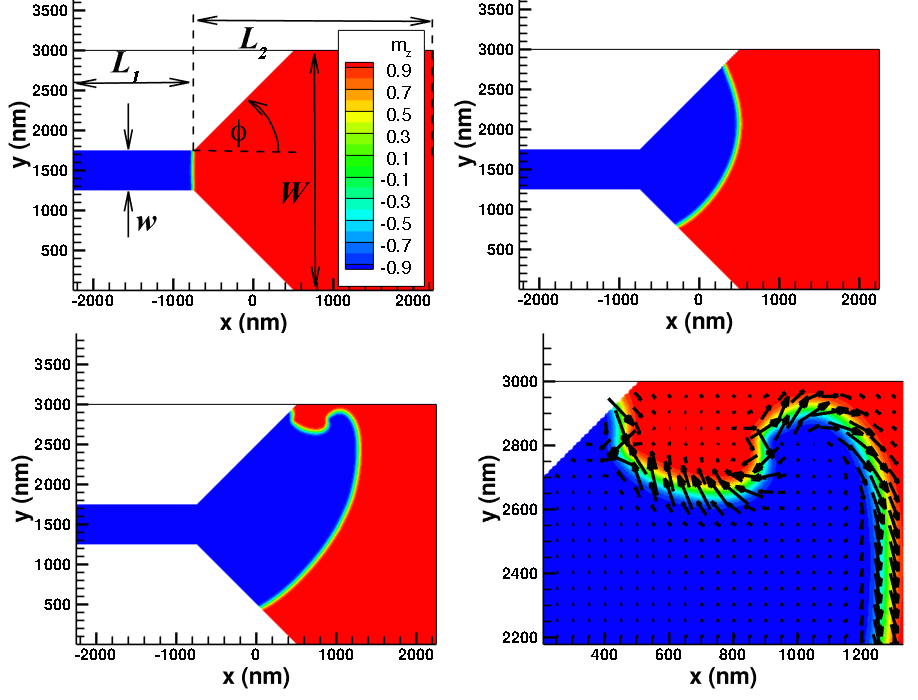}
\caption{(Color online) The evolution of the domain wall under an applied voltage of 0.5~V. Upper left panel: Initial position of a domain wall at the beginning of the flare. The color coding shows the out-of-plane $\hat z$-component of the magnetization; the DMI interaction favors in-plane magnetization pointing from negative (blue) $\hat z$-component to positive (red). Upper right panel: Domain wall position after 8~ns. The domain wall has started to deform slightly at the
upper boundary. Lower left panel: The domain wall after 12~ns. The instability has started to deform the domain wall. Lower right panel: Close-up of the domain wall near the upper boundary. The competition between exchange, which tries to minimize the domain wall, DMI, which tries to make the magnetization in the domain wall point from blue to red 
, the spin-Hall field, which approximately points downward (in the $-\hat y$ direction), and the DMI boundary condition builds up energy density at the boundary. This leads to an instability and a bubble breaks off to release energy.}
\label{fig:figure2}
\end{figure}

We first discuss the domain wall motion and instability for low current densities, with an applied voltage of $0.5$~V and the current density in the
constriction about $10$~MA/cm$^2$. As can be seen in Fig.~\ref{fig:figure2} upper right panel, the domain wall is initially pushed into the flare by the spin-Hall torque.
After some distance, the domain wall starts to deform at the upper edge, and the domain wall becomes unstable and forms a bubble. This bubble
in itself becomes unstable and spawns other bubbles, and so on. 
The initial instability is caused by the competition between spin-Hall torque, exchange, DMI, and DMI boundary conditions\cite{RohartPRB2013} . Figure~\ref{fig:figure2} lower right panel
shows a close-up of the domain wall at the edge, with the arrows denoting the in-plane magnetization direction. In this figure,
the spin-Hall torque is predominately directed downwards (-$\hat y$-direction), while the DMI boundary condition imposes a nonzero outward
magnetization component perpendicular to the boundary of the device. At the same time, the DMI exerts a torque on the domain wall
that tries to direct the magnetization in the domain wall to point from the region with magnetization in the $-\hat z$-direction to
the region where the magnetization is in the $+\hat z$-direction (from blue to red in Fig.~\ref{fig:figure2}). 

\begin{figure}
\includegraphics[width=3.5in]{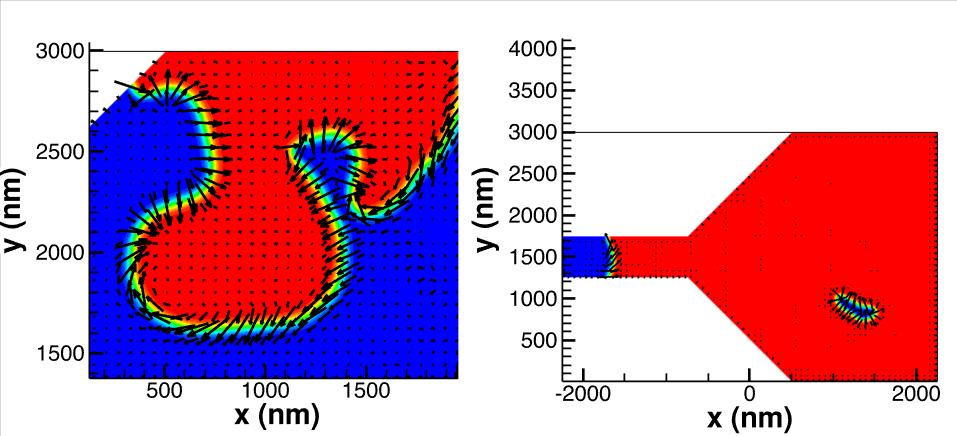}
\caption{(Color online) Left panel: After the initial instability, a bubble moves downward. The magnetization in the lower domain wall of the bubble is in a high-energy state for the DMI interaction as the magnetization here points from red ($+\hat z$ domain) to blue $-\hat z$ domain). This eventually leads to instabilities along this domain wall, and the bubble breaks up into smaller bubbles. Right panel: end magnetization after the system has relaxed for 40~ns after the applied voltage was removed. }
\label{fig:figure5}
\end{figure}

As the domain wall moves into the flare, energy density 
builds up near the boundary because of the competing torques, and a bubble region breaks off and moves downward. This bubble
is not a chiral bubble as it has two Bloch points along the domain wall, with the magnetization pointing from the $+\hat z$-magnetization inside the bubble towards the $-\hat z$ magnetization 
outside the bubble at the lower boundary of the bubble (Fig.~\ref{fig:figure5}).
This magnetization configuration is also unstable as the lower domain wall
is in a high-energy configuration for the DMI, and the lower domain wall eventually breaks up into smaller pieces or bubbles. This process
continues dynamically  and the magnetization breaks up in a rather chaotic structure with many bubbles of different sizes in the flare region of the device so long as the 
voltage (and the current) is applied. This instability is different from what would be observed with a large DMI coupling, $D/D_c > 1$. In that case, at low voltages the domain wall would be pinched off to form a skyrmion bubble with a topological charge of $\pm1$ as it enters the flare, with the main driving forces the DMI interaction and the inhomogeneous spin-Hall torque\cite{Jiang2015,YanzhouNatCom2014}. This is in contrast with the low-DMI skyrmions studied here, which first form bubbles with non-chiral domain walls and that typically contain Bloch points. 

When the applied voltage is removed after 30 -- 40~ns the current and the spin-Hall torque disappear, and the system starts to relax. Depending on the system parameters, the bubble regions start
to coalesce and the domain walls become chiral by emitting spin waves. The regions with $+\hat z$ magnetization grow because of the applied
external field, but some isolated bubbles with $-\hat z$ magnetization are left behind; these bubbles are isolated skyrmion bubbles. The right panel of Fig.~\ref{fig:figure5} shows the magnetization after relaxing for 40 ns. The regions with negative $z$-component of the magnetization has coalesced and retracted into the constriction, but a skyrmion bubble remains in the wider region. The movie S1 in the Supplemental Material at [URL] shows the evolution of the system first under the 30~ns of applied voltage of 0.5~V followed by 40~ns relaxation with the voltage removed.

%
If we increase the applied voltage from 0.5~V to $3$~V, the spin-Hall torque 
 is large enough to create some initial textures, both in the constriction as well as in the wider part of the device. 
A domain wall then forms in the narrow part along its length. 
This domain wall is pushed into the flare of the
device near the center of the wide part and then out to the full length of the geometry. In this case, there is no mechanism to produce peristent instabilities, at least not in the modeling we have performed. The domain wall simply persists more or
less in a steady state under the applied voltage. When the voltage is removed, the $-\hat z$ domain shrinks under the applied out-of-plane
field, and 
eventually retracts into the constriction and disappears.
(See movie S2 in the Supplemental Material at [URL]). 

\begin{figure}
\includegraphics[width=3.5in]{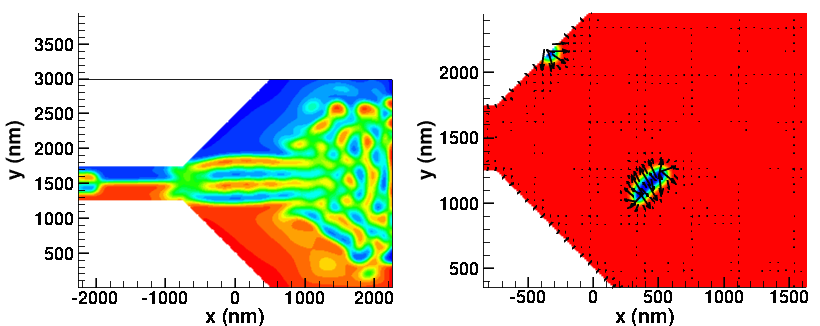}
\caption{(Color online) Left panel: Steady-state spin magnetization under the application of a voltage of $4$~V. The color coding is the out-of-plane magnetization component as in Fig.~\ref{fig:figure2}. Right panel: out-of-plane magnetization component of the device after 40~ns relaxation following 40~ns of an applied voltage of 5~V. 
}
\label{fig:skyrmion_endpoint}
\end{figure}

\begin{figure}
\includegraphics[width=3.5in]{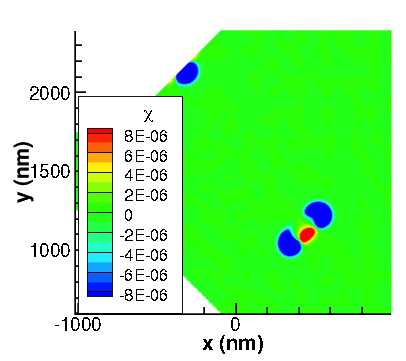}
\caption{(Color online) Skyrmion charge density after a relaxation of 80~ns following a 40~ns applied voltage of 5~V.}
\label{fig:skyrmion_charge}
\end{figure}

If we further increase the voltage to $4$~V or $5$~V a
different instability occurs. Now the spin-Hall torque is strong enough to form a domain wall-like texture in the constriction and also to rotate the magnetization in-plane in the flare and wide regions of the device. This formation grows unstable at the point along the upper edge where the device flares, again because of competing spin-Hall and DMI torques. This leads to the formation of a growing region along the upper edge with the magnetization reversed and pointing along the $-\hat z$ direction, and eventually to the formation of almost a steady state modulated
structure with a distinct spatial period so long as the voltage is applied 
(Fig.~\ref{fig:skyrmion_endpoint} left panel).
The integrated topological charge of the system remains zero. When the voltage is removed, this structure too relaxes with regions
coalescing and forming chiral domain walls. The regions with $+\hat z$-magnetization grows but some chiral bubbles remain and appear to be
stable. Figure~\ref{fig:skyrmion_endpoint} depicts the magnetization in the device after a total of 80~ns; the blue ($-\hat z$-domain) has mostly retracted but left two bubbles in the device, one at the upper boundary and one in the middle of the wider region. (See the movie S3 in the Supplemental Material at [URL].)
The integrated skyrmion charge density of the entire device is here $-1.25$, but the skyrmion charge in a region containing the lower bubble is $-1$. 
This, in addition to the orientation of the magnetization in the bubble domain wall (left panel of Fig.~\ref{fig:skyrmion_endpoint}) conclusively shows that the isolated bubble is a skyrmion bubble. 
During
the application of the 4 or 5~V pulse, the topological charge of the system oscillates around zero with a magnitude of about $0.1$. As the voltage is removed and the system relaxes, the total topological charge of the system oscillates as bubbles coalesce and domain walls are expelled or annihilated. The topologcal charge in the region containing the single bubble increases rapidly in magnitude to unity as the bubble forms and stabilizes.
Figure~\ref{fig:skyrmion_charge} shows the skyrmion charge density $\chi=\frac{1}{4\pi}{\hat m}\cdot\left[\partial_x{\hat m}
\times\partial_y{\hat m}\right]$ after a relaxation of 80~ns following a 40~ns applied voltage of 5~V. The charge density integrated over the whole device is -1.5; the charge density integrated over ar region containing the bubble in the lower right is -1.
 In general, the skyrmion bubbles left behind as the magnetization relaxes are dynamical in that they
 are typically slightly irregular in shape and exibit domain wall oscillations for the duration of the simulations that we have performed. Of course, even with a small damping ($\alpha=0.02$), energy is eventually dissipated and the oscillations will cease. With no energy pumped into the system, the bubbles are topologically protected: the topological charge (here $q=-1$) is conserved and the isolated bubbles remain. 
We speculate that imperfections and pinning sites, present in the real, experimental system, pin the bubbles and further stablize them.

\section{Summary and conclusion}
In conclusion, we have investigated the generation of skyrmion bubbles by strongly inhomogeneous spin-Hall torques. The simulations used parameters determined, as much as possible, from the experimental samples in Ref.~\onlinecite{Jiang2015} and were consequently performed in a regime of low DMI coupling. The results
are consistent with experimental observations of the generation of skyrmion bubbles. We identify two distinct mechanisms, one at lower current densities and spin-Hall torques, and one at larger current densities. They both lead to the
generation of skyrmion bubbles through instabilities, and both mechanisms stem from the competition between spin-Hall, DMI, and exchange
torques. The mechanism at larger current densities injects a steady state magnetization texture into the wider part of the device, which
then coalesces and leaves chiral bubbles behind when the voltage, and spin-Hall torque, is removed. 
Interestingly, this second mechanism does not rely on the presence of a domain wall and indeed we recently observed experimentally skyrmion bubble formation at higher currents even by using non-magnetic point contacts for establishing inhomogeneous currents.
Our results show that stable dynamical skyrmion bubbles can form in the low-DMI regime with $D/D_c\approx0.5$. We have performed simulations with a range of exchange and DMI parameters (keeping the external field and the saturation magnetization density fixed) which confirm our results. We note, however, that we have not been able through modeling to obtain stable skyrmion bubbles for $D/D_c\alt0.5$.
The skyrmion bubbles appear to be dynamically stable over long times. Dynamically stabilized skyrmions have been observed previously\cite{YanzhouNatCom2015}, although in that work energy was pumped into the system via spin transfer torque. In contrast, in the systems studied here, there is no energy pumped into the system once the voltage has been turn off. Nevertheless, the skyrmions persist for very long times.

\begin{acknowledgments}
This work was supported by the Department of Energy, Office of Science, Basic Energy Sciences Materials Science
and Engineering Division. We gratefully acknowledge the computing resources provided on Blues, a high-performance computing cluster operated by the Laboratory Computing Resource Center at Argonne National Laboratory.
\end{acknowledgments}

\bibliographystyle{aipnum4-1}

\bibliography{STO}





\end{document}